\documentclass[prl,amsfonts,amsmath,twocolumn,nofootinbib]{revtex4}

\usepackage{color}
\usepackage{graphicx} 
\usepackage{epstopdf}%
\usepackage{mathptmx}   % times w/mathptmx fonts for math

\newcommand {\be} {\begin{equation}} 
\newcommand {\ba}{\begin{eqnarray}} 
\newcommand {\ee} {\end{equation}} 
\newcommand{\ea} {\end{eqnarray}}

\renewcommand{\Im}{{\rm Im\,}} 
\renewcommand{\Re}{{\rm Re\,}}

\begin{document}

\title{Gamma-$Z$ box contributions to parity violating elastic $e$-$p$ scattering}

\author{Benjamin C. Rislow}

\author{Carl E.\ Carlson}

\affiliation{Helmholtz Institut Mainz, Johannes Gutenberg-Universit\"at, D-55099 Mainz, Germany}
\affiliation{Department of Physics, College of William and Mary, Williamsburg, VA 23187, USA}

\date{\today}

\begin{abstract}

Parity-violating (PV) elastic electron-proton scattering measures Q-weak for the proton, $Q_W^p$.  To extract $Q_W^p$ from data, all radiative corrections must be well-known.  Recently, disagreement on the $\gamma Z$-box contribution to $Q_W^p$ has prompted the need for further analysis of this term.  Here, we support one choice of a debated factor, go beyond the previously assumed equality of electromagnetic and $\gamma Z$ structure functions, and find an analytic result for one of the $\gamma Z$-box integrals.  Our numerical evaluation of the $\gamma Z$-box is in agreement within errors with previous reports, albeit somewhat larger in central value, and is within the uncertainty requirements of current experiments.  

\end{abstract}

\maketitle

%%%%%%%%%%%%%%%%%%%%%%%%%%%%%%%%%%%%%%%%%%%%

%\section{Introduction}			\label{sec:one}

%%%%%%%%%%%%%%%%%%%%%%%%%%%%%%%%%%%%%%%%%%%%

Parity-violating (PV) elastic electron-proton scattering measures $Q$-weak for the proton, $Q_W^p$, which is the $Z$-current Dirac form factor for the proton.  This form factor has been measured at a number of energy scales, and part of the motivation for the PV experiment is to check its evolution against predictions of the standard model.  Deviations from the predictions could be a signal of new physics, that is, of currently unknown terms in the Lagrangian.  Another motive is to check the consistency of $Q_W^p$ measured by NuTeV at low momentum transfer~\cite{Zeller:2001hh} versus measurements using other processes,  although this motive may today be weaker due to the recognition~\cite{Bentz:2009yy} of charge symmetry violating effects upon the NuTeV and other experiments with targets using large nuclei.

Knowing all corrections is important to accurately obtaining $Q_W^p$ from the $e$-$p$ parity violating asymmetry.  Corrections from, among other sources, $\gamma\gamma$, $WW$, and $ZZ$ boxes have been well considered.  A surprise came when Gorchtein and Horowitz~\cite{Gorchtein:2008px} evaluated the inelastic corrections to the $\gamma Z$ box at zero overall momentum transfer (inelastic meaning the hadron state between the $\gamma$ and $Z$ connections is not a proton), using a dispersive method that connected the box evaluation to the inelastic structure functions.  Their result was unexpectedly large and of of uncertain robustness.  Sibirtsev \textit{et al.}~\cite{Sibirtsev:2010zg} subsequently reevaluated the $\gamma Z$ box, finding even larger corrections but importantly asserting that the uncertainty in the corrections was safely below the projected uncertainty in the experimental result.  This is in line with the conclusions in~\cite{Gorchtein:2010ve}.

Because of the importance of the consequences and of the differences in the two results, including an overall factor "2" in one of the main formulas,  we present another visit to this subject.  We corroborate the factor "2" as given by Sibirtsev \textit{et al.} and give numerical evaluations with at least partly different input that leads to results slightly larger but compatible within uncertainty limits compared to Sibirtsev \textit{et al}.  We also show a useful technical advance not mentioned in~\cite{Gorchtein:2008px} or~\cite{Sibirtsev:2010zg}, that one of the triple integrals required to obtain the answer can be analytically done, leading to an easier numerical evaluation of the final result.  Our results are still at zero momentum transfer;  a partonic calculation of the $\gamma Z$ box valid at high momentum transfer can be found in~\cite{Chen:2009mza}.

%We review the calculation of the $\gamma Z$ box in the next section, give our numerical evaluation of the box in Sec.~\ref{sec:numerical}, and conclude with a discussion in Sec.~\ref{sec:end}.

%%%%%%%%%%%%%%%%%%%%%%%%%%%%%%%

%\section{Calculation of $\gamma Z$ box diagrams}

%%%%%%%%%%%%%%%%%%%%%%%%%%%%%%%

\begin{figure}[b]
\begin{center}
\includegraphics[width = 3.37 in]{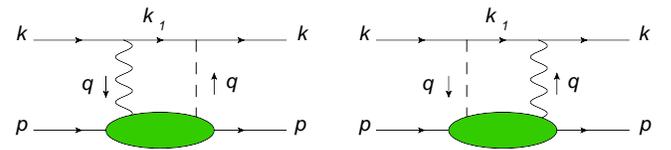}
\caption{The $\gamma$-$Z$ box diagrams.}
\label{fig:mgz}
\end{center}
\end{figure}

\textit{Calculation of $\gamma Z$ box diagrams.}  The quantity $Q_W^p$ is the Z-boson current Dirac form factor of the proton, evaluated at zero momentum transfer.  One can measure it from the parity violating asymmetry in elastic electron-proton scattering,
\be
A_{PV} = \frac{\sigma_R -\sigma_L}{\sigma_R +\sigma_L}	\,,
\ee
where $\sigma_{R,L}$ are cross sections for electron helicities $\lambda = \pm 1/2$ and unpolarized protons.  To lowest order, $A_{PV}$ comes from interference between single $\gamma$ and single $Z$ exchanges, and
\be
A_{PV}^{LO} = \frac{G_F}{4\pi\alpha\sqrt{2}} \, t \, Q_W^{p,LO}	\,,
\ee
where $t$ is the overall momentum transfer, negative for spacelike momentum transfers, and $Q_W^{p,LO} = 1 - 4\sin^2\theta_W(0)$.  With corrections, one has, following~\cite{Erler:2003yk},
\be
						\label{eq:corrections}
Q_W^p = \left( 1+\Delta\rho + \Delta_e \right) \left(Q_W^{p,LO} + \Delta'_e \right)
	+ \square_{WW} + \square_{ZZ} + \Re\square_{\gamma Z} .
\ee
The $WW$ and $ZZ$ box diagrams give $\square_{WW}$ and $\square_{ZZ}$, and are well calculated perturbatively.  The $\gamma Z$ box diagrams, Fig.~\ref{fig:mgz}, involve low momentum scales where perturbation theory is not reliable for the hadronic part of the diagram.  Gorchtein and Horowitz~\cite{Gorchtein:2008px}, calculating only contributions from the inelastic intermediate states (elastic contributions have been considered in~\cite{Zhou:2007hr,Zhou:2009nf,Tjon:2007wx,Tjon:2009hf}), showed how to dispersively relate the $\gamma Z$ box at $t=0$ to hadronic structure functions.  With some approximations, they obtained a result that was larger than expected.  Sibirtsev \textit{et al.}~\cite{Sibirtsev:2010zg} improved the calculation, obtaining in fact a somewhat larger result but with tighter uncertainty limits.

%%%%%%%%%%%%%

%%%%%%%%%%%%%

Demands on the uncertainty limits are set by current and planned experiments.  The Q-weak experiment at JLab aims to measure $Q_W^p$ to about 4\% combined statistical and systematic error at an incoming electron energy of 1.165 GeV and MAMI is discussing an experiment with 180 MeV incoming electron energy measuring $Q_W^p$ to perhaps a part in a thousand.

Theoretically, the quantity $\square_{\gamma Z}$ is obtained from the parity violating part of ${\cal M}_{\gamma Z}$, Fig.~\ref{fig:mgz}, by comparison to the corresponding term in single $Z$-exchange,
\be
\square_{\gamma Z} = 
\frac{{\cal M}_{\gamma Z; \lambda = 1/2} - {\cal M}_{\gamma Z; \lambda = -1/2}}
	{{\cal M}_{Z; \lambda = 1/2} - {\cal M}_{Z; \lambda = -1/2}} \, Q_W^{p,LO} \,.
\ee
(Reference~\cite{Gorchtein:2008px} presents results using $\delta_{\gamma Z} = \square_{\gamma Z}/Q_W^{p,LO}$.)

For the exchange of a Z-boson between an electron with momentum $k$ and proton with momentum $p$, the denominator of $\square_{\gamma Z}$ is
\be
\label{Mzdiff}
{\cal M}_{Z;\lambda=1/2}-{\cal M}_{Z;\lambda=-1/2}=\frac{8}{\sqrt{2}} G_FQ^{p,LO}_W p\cdot k \, g^e_A.
\ee
where $g^e_A=-\frac{1}{2}$.

The calculation of the numerator of $\Re\square_{\gamma z}$ requires the application of the optical theorem.  The imaginary portion of the amplitude for photon, Z-boson exchange is
\begin{align}
\label{imamp}
\Im{\cal M}_{\gamma Z}&=-\frac{1}{2}e^2\left(\frac{g}{2\cos\theta_W}\right)^2\int\frac{d^3\vec{k}_1}{(2\pi)^32E_1}
\frac{ 4\pi L^{\mu \nu}_{\gamma Z}W_{\mu \nu}^{\gamma Z} }{ q^2 (q^2-M^2_Z) }
\end{align}
with 
\begin{align}
\label{Lmunu}
L^{\mu \nu}_{\gamma Z} &= 2(g^e_V-(2\lambda)g^e_A)
			\nonumber \\
&\times (k^\mu_1k^\nu+k^\nu_1k^\mu-k_1\cdot kg^{\mu \nu}+i(2\lambda) \epsilon^{\mu \nu \alpha \beta}k_{\alpha}k_{1\beta}),
\end{align}
\begin{align}
\label{Wmunu}
W_{\mu \nu}^{\gamma Z}&=\frac{1}{4\pi}\int d^4 \eta e^{iq\eta}\langle ps\left|J_{Z\mu}(\eta)J_{\gamma\nu}(0)+J_{\gamma\mu}(\eta)J_{Z\nu}(0\right|ps\rangle
\nonumber \\
&=\big[ \left(-g_{\mu \nu}+\frac{q_\mu q_\nu}{q^2} \right) F^{\gamma Z}_1(x,Q^2)+\frac{p_\mu p_\nu}{p\cdot q}F^{\gamma Z}_2(x,Q^2)
       \nonumber\\
& -i\epsilon_{\mu\nu\alpha\beta}\frac{q^\alpha p^\beta}{2p\cdot q}F^{\gamma Z}_3(x,Q^2) \big]    .
\end{align}
Here,  $k_1$ is the intermediate 4-momenta of the electron,  $Q^2=-q^2$,  and $g^e_V=-\frac{1}{2}+2\text{sin}^2\theta_W$.  
One obtains
\begin{align}
\label{Mgammazdiff}
&\Im \left( {\cal M}_{\gamma Z ;\lambda=1/2}
		-{\cal M}_{\gamma Z ;\lambda=-1/2} \right)
			\nonumber \\
&\hskip 6 mm = \frac{16\pi}{\sqrt{2}}G_Fe^2 \int \frac{d^3\vec{k}_1}{(2\pi)^32E_1}\frac{1}{1 + Q^2/M^2_Z}
\nonumber \\
& \times \left[g^e_A\left(F^{\gamma Z}_1(x,Q^2)+AF^{\gamma Z}_2(x,Q^2)\right)+g^e_V B \, F^{\gamma Z}_3(x,Q^2)\right]
\end{align}
where
\be
\label{A}
A= \frac{2p\cdot k_1 p\cdot k}{Q^2 p\cdot q}-\frac{p^2}{2p\cdot q}
\quad {\rm and} \quad
B=\frac{p\cdot k+p\cdot k_1}{2p\cdot q}.
\ee

Upon changing integration variables, $\Im\square_{\gamma Z}$ becomes
\begin{align}
\label{imdelta3}
\Im \square^V_{\gamma Z}(E) &= \frac{\alpha}{(2ME)^2}
\int^s_{W^2_\pi} dW^2 
				\nonumber \\
& \times \int^{Q^2_{max}}_0 dQ^2 \frac{ F^{\gamma Z}_1(x,Q^2)+AF^{\gamma Z}_2(x,Q^2)}{1 + Q^2/M^2_Z }	\,,
\end{align}
where $W_\pi^2=(M+m_\pi)^2$, $m_\pi$ is the mass of the pion, and $Q^2_{max}=(s-M^2)(s-W^2)/s$.  The $F_3^{\gamma Z}$ structure function has been dropped because the ratio $g^e_V / g^e_A \approx 0$.  The remaining term is labeled by a superscript "$V$" to indicated its association with the vector part of the $Z$-boson current.  This expression agrees with the result reported in ~\cite{Sibirtsev:2010zg}.

The real part, $\Re \square_{\gamma Z}$, is given by the dispersion relation,
\be
\label{redelta}
\Re \square^V_{\gamma Z}(E)=\frac{2E}{\pi} \int^\infty_{\nu_\pi}\frac{dE'}{E'^2-E^2}\Im \square^V_{\gamma Z}(E') 
\ee
where $\nu_\pi=(W^2_\pi-M^2)/2M$.

Rewriting Eq.(\ref{A}) in the current context as 
\be
\label{A2}
A=\frac{(2ME')^2+2ME'(M^2-W^2-Q^2)-M^2Q^2}{Q^2(-M^2+W^2+Q^2)}  \,,
\ee
we notice that the $E'$ dependent terms can be separated from those dependent on $Q^2$ and $W^2$.  The $E'$ integrands can be evaluated analytically if the order of integration is switched.  Sliding the energy integration to the inside changes all three integration bounds.  The $W^2$ integral is now evaluated from $W_\pi^2$ to $\infty$, $Q^2$ from $0$ to $\infty$, and $E'$ from $E_{min}=\frac{1}{4M}[W^2-M^2+Q^2+((W^2-M^2+Q^2)^2+4M^2Q^2)^{1/2}]$ to $\infty$.  After evaluating the $E'$ integrals, $\Re\square^V_{\gamma Z}$ becomes
\begin{align}
\label{redelta3}
&\Re \square^V_{\gamma Z}(E)=-\frac{\alpha}{2\pi M^2E}
\int^\infty_{W^2_\pi}dW^2\int^\infty_0  \frac{dQ^2}{1+Q^2/M_Z^2}
	\nonumber \\
&  \times \bigg\{ \left[\frac{1}{E_{min}}+\frac{1}{2E}
\ln\left(\frac{|E_{min}-E|}{E_{min}+E}\right)\right]
\left(  F_1-\frac{M^2 F_2^{\gamma Z}}{W^2-M^2+Q^2}  \right)
	\nonumber \\
& \quad -  \ln \left(\frac{|E^2_{min}-E^2|}{E^2_{min}}\right)
\frac{ M F_2^{\gamma Z}}{Q^2}
	\nonumber \\
& \quad   + \ln\left(\frac{|E_{min}-E|}{E_{min}+E}\right)
\frac{ 2 M^2 E F_2^{\gamma Z}}{Q^2  (W^2-M^2+Q^2)}  \bigg\}.
\end{align}

%%%%%%%%%%%%%%%%%%%%%%%%%%

%\section{Evaluation of $\Re\square_{\gamma Z}^V$}   \label{sec:numerical}

%%%%%%%%%%%%%%%%%%%%%%%%%%

\textit{Evaluation of $\Re\square_{\gamma Z}^V$.}  Experimental data do not exist for $F_{1,2}^{\gamma Z}$.  In the scaling region, high $Q^2$ and high $W$, there are separated parton distributions~\cite{Lai:2010vv,Martin:2009iq} and one gets $F_{1,2}^{\gamma Z}$ using
\be
F_2^{\gamma Z} = x \sum_{q, \bar q} 2 e_q g_q^V f_q(x,Q^2) \,,
\ee
similar to the purely electromagnetic $F_{1,2}^{\gamma \gamma}$ where $2 e_q g_q^V \to e_q^2$.  However, one expects and can verify that the bulk of the support for the $\gamma Z$ box comes from the resonance region and from lower $Q^2$.  In order to proceed, earlier work accepted~\cite{Gorchtein:2008px,Sibirtsev:2010zg} the equality $F_{1,2}^{\gamma Z} = F_{1,2}^{\gamma \gamma}$, which can be shown to be approximately true in certain regions and certain limits.  We will investigate the equality and improve upon it.

Our numerical evaluation of $\Re\square_{\gamma Z}^V$  uses the Christy-Bosted fits~\cite{Christy:2007ve} in the resonance region ($W < 2.5$ GeV), the Capella \textit{et al.} fits in the high-energy low-$Q^2$ region ($W > 2.5$ GeV and $Q^2 < 5$ GeV$^2$), both of these with some modification, and used the CTEQ parton distributions CT10.00~\cite{Lai:2010vv} in the scaling region ($W > 2.5$ GeV and $Q^2 > 5$ GeV$^2$).  

For resonance photoproduction and electroproduction, the parton model (e.g.,~\cite{Close:1979bt}) shows how each amplitude depends on the quark charges.  It is useful to note that later analysis indicated that two-quark operators play a small role in photoproduction amplitudes~\cite{Carlson:1998si}.   The charges then can be changed to the $Z$-boson vector coupling parameters $g_V^q$ to compare resonance contributions in $F_{1,2}^{\gamma Z}$ and $F_{1,2}^{\gamma \gamma}$.  For any isospin-3/2 resonance, the result is just a multiplication by $(1+Q_W^{p,LO})$,  since only the $\Delta I = 1$ currents contribute~\cite{Sibirtsev:2010zg}.   

Other resonances are more complicated.  For example, for the $D_{13}(1520)$, the $A_{3/2}$ amplitude also scales like $(1+Q_W^{p,LO})$, but the $A_{1/2}$ amplitude has two contributions, one sharing a matrix element with $A_{3/2}$ and one which will be multiplied by $(1/3 + Q_W^{p,LO})$.   There is extra phenomenological information, that the $A_{3/2}$ dominates in photoproduction and that there is a rapid transition to the high $Q^2$ dominance of the $A_{1/2}$ expected from hadron helicity conservation~\cite{Brodsky:1981kj,Carlson:1985mm}.  This give enough information to modify the Christy-Bosted $D_{13}$ contribution for the $\gamma Z$ in a $Q^2$ dependent fashion.  As a remark, the average $Q^2$ within the integrals for incoming energies in the JLab range is only about $0.4$ GeV$^2$.   Similar considerations apply to the $F_{15}(1690)$, although now the multiplication factors are $Q_W^{p,LO}$ for the $A_{3/2}$ and $(2/3 + Q_W^{p,LO})$ for the other amplitude, so the reduction from the purely electromagnetic case is quite noticeable.  The modification of the resonant part of the Christy-Bosted fit is thus straightforwardly done, and gives a resonant contribution to $F_{1,2}^{\gamma Z}$ about 9\% smaller than to $F_{1,2}^{\gamma \gamma}$.  The Christy-Bosted fits come within 3\% of nearly all the data points, and the points themselves have comparable (mostly systematic) error. We allow some margin, assigning a 10\% uncertainty in this part of the calculation.   

An additional note is that the amplitude for electromagnetic excitation of a proton to a state with quark spin-3/2 is proportional to $(e_u+2e_d)$ (this is the Moorhouse selection rule~\cite{Moorhouse:1966jn}),  which is not zero when turned into its $Z$-current analog.  However, this excitation seems small also for a neutron target, so we do not consider it further.

\begin{figure}[t]
\begin{center}
\includegraphics[width = 84 mm]{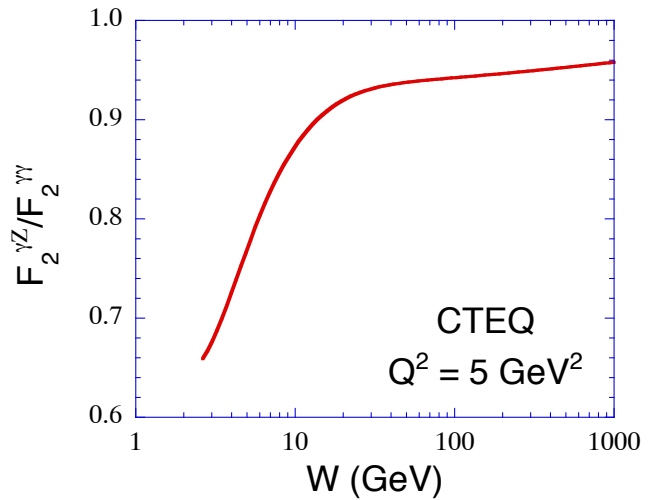}
\caption{The ratio $F_{2}^{\gamma Z}/F_{2}^{\gamma \gamma}$ vs.~$W$ obtained from the CTEQ parton distribution functions at fixed $Q^2 = 5$ GeV.
}
\label{fig:f2equalityW}
\end{center}
\end{figure}

The resonance region fit includes a smooth background non-resonant part,  which one can think of as scattering off collections of quarks with scant final state interactions.    In a full $SU_f(3)$ limit, where all light quarks are equally likely and which may be pertinent in a high-energy $x Q^2/(2M\nu)\to 0$ limit, one has $F_{1,2}^{\gamma Z}/F_{1,2}^{\gamma \gamma} = 1 + Q_W^{p,LO}$.  In a valence quark limit with SU(6) wave functions, one gets $(2/3 + Q_W^{p,LO})$ for the same ratio.  The latter is better at high-$x$ and the former is better at low-$x$ and we take the mean, and use the extremes to set our uncertainty estimate.   One can examine the $F_{1,2}^{\gamma Z}/F_{1,2}^{\gamma \gamma}$ ratio in the scaling region, and the result along one boundary of the CTEQ region is shown in Fig.~\ref{fig:f2equalityW}.   The value at the low $W$ end is in agreement with out high-$x$ expectation for the background in the resonance region.  The rest of this CTEQ-based curve is at lower-$x$ and the $F_{1,2}^{\gamma Z}$ and $F_{1,2}^{\gamma \gamma}$ structure functions are nearly within 5\% of equality for much of the range.  This also marks what we may expect at the upper end of the Capella \textit{et al.} region, with expectation of closer equality as $Q^2$ further decreases.  We estimate the modification of the Capella \textit{et al.} fit for the present case by multiplying it by a $W$ dependent function which is the mean of unity and this boundary curve, and take the extremes to estimate the uncertainty.

%%%%%%%%%%%%%%%
%%%%%%%%%%%%%%%

%%%%%%%%%%%%%%%
\begin{figure}[b]
\begin{center}
\includegraphics[width = 75 mm]{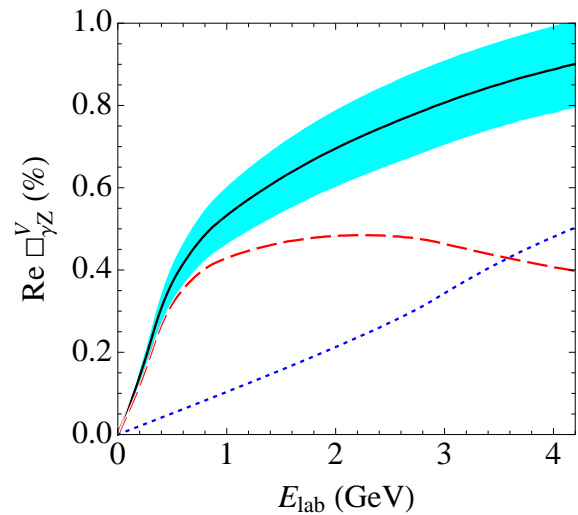}
\caption{Plot of $\Re \square_{\gamma Z}^V$ vs. incoming electron lab energy.   The dashed red line gives the resonance region contribution using the Christy and Bosted~\cite{Christy:2007ve} structure function fit; the dotted blue curve gives the non-resonance region contribution using Capella \textit{et al.}~\cite{Capella:1994cr} and CTEQ \textit{et al.}~\cite{Lai:2010vv}.  The solid black curve gives the total, with an error band indicated.   %and using Golec-Biernat and W\"usthoff~\cite{GolecBiernat:1998js} gives the dashed red curve.
}
\label{fig:boxvse}
\end{center}
\end{figure}
%%%%%%%%%%%%%%%

Our numerical results for $\Re\square_{\gamma Z}^V$ as a function incoming electron energy  are shown in Fig.~\ref{fig:boxvse}.  We also show the separate results from the resonance region and from above the resonance region, and show uncertainty limits for the total.     For the JLab Q-weak experiment, $E_{lab} = 1.165$ GeV, and 
\be
\Re\square_{\gamma Z}^V (1.165 \rm{\ GeV})= 0.0057 \pm 0.0009 \,.
\ee
This agrees within uncertainty limits with the Sibirtsev \textit{et al.} result  $0.0047^{+0.0011}_{-0.0004}$.  

For information, at the JLab energy, the high $Q^2$ above the resonance region contribution from CTEQ gives 0.00019 of the total;   this is about 0.00004 lower would be gotten by simply extending Capella \textit{et al.} to the high $Q^2$ region.  Also, the longitudinal part of the structure functions contribute only about 0.0007 to the above result, roughly evenly split among resonances in the resonance region, non-resonant background with $W<2.5$ GeV, and contributions where $W>2.5$ GeV.

\begin{figure}[t]
\begin{center}
\includegraphics[width = 80 mm]{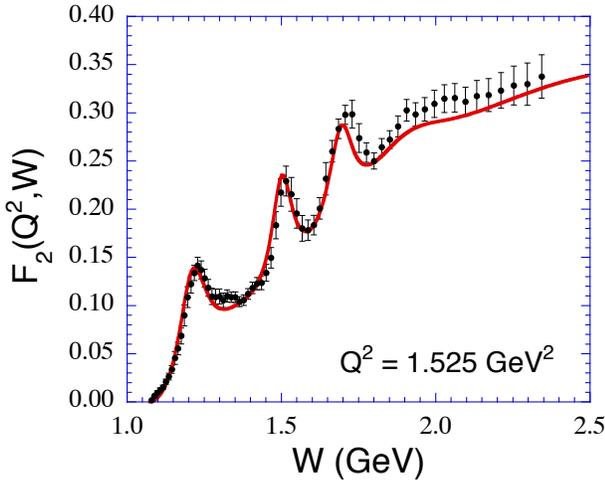}
\caption{Plot of $F_2(Q^2,W)$ vs. $W$ using~\cite{Christy:2007ve} at $Q^2 = 1.525$ GeV$^2$.  The data is from JLab~\cite{Osipenko:2003bu,Osipenko:2003ua}.
}
\label{fig:bc}
\end{center}
\vglue -14 pt

\end{figure}

Though we agree with the Sibirtsev \textit{et al.} result, part of the agreement is due to the reduction in our result from analyzing the $F_{2}^{\gamma Z} = F_{2}^{\gamma \gamma}$ relation.  Had we used the equality everywhere but the scaling region, our result would have been $0.00065$ higher.  
One difference between us is that in the resonance region, we used the Christy-Bosted fit~\cite{Christy:2007ve}, which represents the data to 3\% or better over almost the entire applicable range.  By way of examples, Christy and Bosted give plots of cross section vs. $W$ at a number of incoming energies and angles.  Sibirtsev \textit{et al.} used their own dedicated resonance region fits, and also fit the data well, as seen in their plots of $F_2$ vs. $W$ at several fixed $Q^2$'s~\cite{Sibirtsev:2010zg}.  To facilitate direct comparison, Fig.~\ref{fig:bc} here shows the Christy-Bosted $F_2$ vs. $W$ at a typical $Q^2$.  

%%%%%%%%%%%%%%%

%%%%%%%%%%%%%%%

%%%%%%%%%%%%%%%%%%%%%%%%%%

%\section{Discussion}				\label{sec:end}

%%%%%%%%%%%%%%%%%%%%%%%%%%

\textit{Discussion.}     Using the $Q^2=0$ value $\sin^2\theta_W(0) = 0.23867(16)$~\cite{Erler:2004in} then $Q_W^{p,LO} = 0.04532 (64)$ and with the corrections listed in Eq.~(\ref{eq:corrections}), $Q_W^{p}$ somewhat exceeds 0.07.  A 4\% measurement of this number requires an absolute accuracy of about $0.0028$,  so the $\Re \square_{\gamma Z}^V$ correction needs to be known more accurately than, say, 30\%.  This we believe is the case.

For the future, discussions of a PV experiment at lower energy~\cite{mami} are partly prompted by the smaller expected $\gamma Z$ box correction.  For 180 MeV we obtain
\be
\Re\square_{\gamma Z}^V(180 \rm{\ MeV}) = 0.00125 \pm 0.00018 \,.
\ee
However,  the goal is now a part in a thousand measurement of $Q_W^p$, so the uncertainty requirement is about 0.00006.  This remains a challenge, and will require further accurate fits in the resonance region, further thinking about the validity of the $F_{1,2}^{\gamma Z} = F_{1,2}^{\gamma \gamma}$ approximation, and considering the $\Re\square_{\gamma Z}^A$ term. 

We conclude by restating that the $\gamma Z$ box contribution is known well enough for current  experiments, but that more accurate determinations will be needed within several years.

%%%%%%%%%%%%%%%%%%%%%%%%%%%%%%%

\begin{acknowledgments}

We thank the National Science Foundation for support under Grant PHY-0855618, thank the Helmholtz Institute at the Johannes Gutenberg-Universit\"at in Mainz and the Helsinki Institute for Physics for their hospitality, and thank Peter Blunden, Wally Melnitchouk, and Marc Vanderhaeghen for useful comments.

\end{acknowledgments}

  \vskip -15pt

%%%%%%%%%%%%%%%%%%%%%%%%%%%%%%%

\bibliography{gazbox}

\newpage

\end{document}